\documentclass[%
 reprint,
 amsmath,amssymb,
 aps,
]{revtex4-1}

\usepackage{graphicx}
\usepackage{dcolumn}
\usepackage{bm}

\newcommand{\bra}[1]{\langle #1|}
\newcommand{\ket}[1]{|#1\rangle}

\newcommand{\braket}[2]{\langle #1|#2\rangle}

\begin{document}

\preprint{APS/123-QED}

\title{How Kirkwood and Probability Distributions Differ: A Coxian Perspective}



\author{Kevin Vanslette}
\email{kvanslette@albany.edu}
\affiliation{
 Department of Physics, University at Albany (SUNY),
 Albany, NY 12222, USA.
}%
%

\date{\today}

\begin{abstract}
Kolmogorov's first axiom of probability is probability takes values between 0 and 1; however, in Cox's derivation of probability having a maximum value of unity is arbitrary since he derives probability as a tool to rank degrees of plausibility. Probability can then be used to make inferences in instances of incomplete information, which is the foundation of Baysian probability theory. This article formulates a rule, which if obeyed, allows probability to take complex values and still be consistent with the interpretation of probability theory as being a tool to rank plausibility. It is then shown that Kirkwood distributions and the conditional complex probability distributions proposed by Hofmann do not obey this rule and therefore cannot rank plausibility. Not only do these quasiprobability distributions relax Kolmogorov's first axiom of probability, they also are void of the defining property of a probability distribution from a Coxian and Baysian perspective - they lack the ability to rank plausibility.
\end{abstract}

\maketitle


\section{\label{sec:level1}Introduction}

Probability theory as derived by Cox (and later Jaynes and Caticha) is a tool an observer can use to rank degrees of plausibility in instances of incomplete information \cite{Cox,Jaynesbook,book}. Accordingly, scientists can use probability to make useful inferences quantified by changes in a probability distribution when new information is attained. Probability functions have a few mathematical properties: they are normalized, $0\leq p(a)\leq 1$, obey a sum and product rule, and marginalize; but arguably the most useful statement in probability theory is Bayes' Rule because it forms the basis of inference in probability theory. Bayes' Rule is represented by,
\begin{eqnarray}
p(a)\rightarrow p'(a)\equiv p(a|b)=p(a)\frac{p(b|a)}{p(b)},
\end{eqnarray}
which quantifies a new probability distribution for ``$a$" once ``$b$" is learned to be true causing the distribution for ``$a$" to change by factors $|p'(a)-p(a)|\geq 0$. a general probability function $p(a|b)$ is a unique function that accomplishes all of the goals above up to arbitrary regraduation (or rescaling) $p(a|b)\equiv \xi([a|b])$, which forces $0\leq p(a|b)\leq 1$ rather than taking arbitrary values $0=v_F\leq [a|b]\leq v_T$ on the real axis. Probability takes real values so one can rank degrees of plausibility from not plausible, $v_F$, to absolute certainty, $v_T$, which inevitably allows probability to be used as a tool for reasoning, inference, and measurement \cite{Cox,Jaynesbook,book}. 

The renewed interest in Weak Values $A_w=\frac{\bra{b}\hat{A}\ket{a}}{\braket{b}{a}}$ \cite{aav,duck,Dressel} has revitalized discussion of Kirkwood distributions $p_K(a,b)=\braket{b}{a}\bra{a}\hat{\rho}\ket{b}$ \cite{Kirkwood,Johansen} and given rise to notions of a complex conditional probability $p_K(m|a,b)=\frac{\braket{b}{m}\braket{m}{a}}{\braket{b}{a}}$ (a conditional Kirkwood distribution) \cite{Hofmann1,Hofmann2,Hofmann3,Sasaki} by Hofmann. Because Weak Values are in principle complex numbers, the measurement (or perhaps more appropriately the inference) of them through post selection $\bra{b}$ of a prepared state $\ket{a}$ has lead to a surplus of obscure and seemingly paradoxical results \cite{Dressel,Lundeen1,cat}.  In the literature, Kirkwood distributions are often likened to probability distributions because they marginalize correctly $p(a)=\sum_bp_K(a,b)=\sum_b \braket{b}{a}\bra{a}\hat{\rho}\ket{b}=\mathtt{Tr}(\hat{\rho}\ket{a}\bra{a})$ and $p(b)=\sum_ap_K(a,b)=\sum_a \braket{b}{a}\bra{a}\hat{\rho}\ket{b}=\mathtt{Tr}(\hat{\rho}\ket{b}\bra{b})$ with the disclaimer that in-fact they are quasiprobability distributions because they take complex values thereby relaxing Kolmogorov's first axiom of probability, $0\leq p(a|b)\leq 1$.

Because Cox and Kolmogorov derive probability theory independently of one another, we can evaluate how Kirkwood distributions and probability distributions differ from a Coxian perspective and obtain additional insight. The extent that Coxian probability theory is compatible with probability functions/theories taking complex values will be shown in the next section. This is then used to show that Kirkwood distributions and complex conditional probability distributions in the sense of Hofmann are unlike and incompatible with standard probability distributions because transitivity is not preserved and therefore Kirkwood distributions cannot rank plausibility. 
\section{\label{sec:level2}The extent that probability can be complex and consistent}

In Coxian derivations of probability theory  \cite{Cox,Jaynesbook,book} one requires the transitive property to rank degrees of plausibility: if $a$ is more plausible than $b$ and $b$ is more plausible then $c$ then $a$ must be more plausible than $c$. If we let $[a|b]$ be the plausibility function of $a$ given $b$ (will eventually be $p(a|b)$ in standard notion) then the transitive property of the plausibility function can be satisfied if $[a|b]$ maps the space of propositions $a\in A$ to the real number line $r\in \mathcal{R}$. To be consistent, the plausibility function has the two extremal values imposed by transitivity: the plausibility of $a$ given $a$ is $[a|a]=v_T$ is true and not-$a$ given $a$, $[\tilde{a}|a]=v_F$, is completely implausible and $v_T > v_F$. Because $v_T\equiv[a \vee \tilde{a}|a]=v_T+v_F$, where $\vee$ is the ``or" logic proposition, we have in general $v_F=0$ giving in principle a lower bound. 

To explore in what sense a probability distribution can be complex and still agree with standard probability theory, let the plausibility function map $a\in A$ to a point $z'=[a|b]$ in the complex plane $\mathbb{C}$. Because in general the inequality $z_1 \leq z_2$ between any two arbitrary complex numbers $z_1,z_2\in \mathbb{C}$ is poorly defined there is no notion (known to the author) of monotonicity in the complex plane. We therefore define a ``monotonic curve" $c'$ to be an arbitrary curve in the complex plane (piece-wise or otherwise) which can be ordered by applying an order preserving transformations $\mathcal{O}$ which maps $c'$ onto the real axis. The trivial examples are if $c'$ lies along the real axis in $\mathbb{C}$ such that $\mathcal{O}=1$ or if $c'$ is constrained to the imaginary axis between $0$ and $i$ then $0\leq [a|b]\leq i$ and therefore $\mathcal{O}=e^{-i\pi/2}=-i$.  In the second case our probability distribution would have the feature that $\sum_a [a|b]\equiv i=v_T$ but this is by no means \emph{physical} - it is simply a reparameterization of what number represents certainty or quantifies a fraction of certainty. To be consistent with probability theory, any ``monotonic curve" $c'$ must be mapped to $\mathcal{R}$ by $\mathcal{O}$ where inferences can be made (probability updated, added, or multiplied) and then it may be mapped back to $c'$ by $\mathcal{O}^{-1}$ to give a complex representation of the plausibility/probability of a proposition. The set of monotonic curves $c'\in\mathbb{C}'$ which can be mapped to $\mathcal{R}$ turns out to be any curve $c'$ in the complex plane which is non-self intersecting (e.g. a spiral with the label $v_F$ and $v_T$ at the endpoints), because any non-self intersecting curve can be `unwound" by $\mathcal{O}$ to unique positions on $\mathcal{R}\geq 0$.  If there exists a point of intersection $z_0'$ on $c'$ then one would have to map it to two points along $\mathcal{R}$ consequently not preserve transitivity and the probability function would lose the interpretation of ``ranking plausibility". Finding the specific operator $\mathcal{O}$ which accomplishes this is for any $c'$ is not the focus of this paper and will not be discussed further. It is also required that $v_T \neq v_F$ for the function $[a|b]$ such that the plausibility has a nonzero measure. If $p_K(a,b)$ and $p_K(m|a,b)$ are consistent with probability theory then must take values along a non self intersecting curve $c'\in \mathbb{C}'$ for arbitrary $a,b,m$, and $v_T,v_F$ must be defined. 
\section{\label{sec:level3} Non-compatibility of Kirkwood distributions with Coxian probability}
Consider the Weak Value of $\ket{\varphi}\bra{\varphi}$ preselected in the $\ket{a}=\ket{\psi_m}=(2\pi)^{-1/2}\int d\varphi \,e^{im\varphi}\ket{\varphi}$ where $m$ is the magnetic/angular momentum quantum number, and post selected in the state $\ket{b}=\sqrt{1-\delta^2}\ket{\psi_n}+\delta\ket{\psi_m}$ (and let $\delta$ be real for simplicity). This Weak Value has the form of a complex (or weak) conditional probability density \cite{Hofmann1,Hofmann2,Hofmann3,Sasaki}, and one finds that
\begin{eqnarray}
p_K(\varphi|a,b)=\frac{\braket{b}{\varphi}\braket{\varphi}{a}}{\braket{b}{a}}=\frac{1}{2\pi}+\frac{\sqrt{1-\delta^2}}{2\pi\delta}e^{i\varphi(m-n)}
\end{eqnarray}
takes values spanning a circle in the complex plane centered at $\frac{1}{2\pi}$ with a radius $\frac{\sqrt{1-\delta^2}}{2\pi\delta}$ which may grow arbitrary large for arbitrarily small $\delta$. Each value of $p_K(\varphi|a,b)$ lies somewhere on a curve $c'_{1}$ (in principle a segment of $c'$) and should indicate a different degree of plausibility for each unique value of $z'(\varphi)=p_K(\varphi|a,b)$, but because no value of $v_F$ or $v_T$ is given for $p_K$, it is impossible in this single case to determine the operation $\mathcal{O}$ which could be used to map $c'$ to $\mathcal{R}$ (we don't know where to cut the circle). Because any value of $p_K(m|a,b)$ must be on the curve $c'$ for the framework to be consistent and because the wavefunction is a Weak Value $p_K(x|p,\Psi)=k\Psi(x)$ \cite{Lundeen1}, which in principle has values spanning another curve $c'_2$ which intersects $c'_1$, it therefore follows that $p_K(m|a,b)$ does not preserve transitivity because $c'$ is a self intersecting curve. The Kirkwood distribution $p_K(a,b)$ and its complex conjugate $p_K(b,a)=p_K^*(a,b)$ are separate distributions, but in principle can output real values $z_0$ in the complex plane, which means the two distributions exist on a self intersecting curve (at the real value). Therefore we arrive at the main result - Kirkwood distributions $p_K(a,b)$ and complex conditional probability distributions $p_K(m|a,b)$ are not consistent with Coxian probability theory because they fail to identify and differentiate degrees of plausibility and cannot be used for inference. It follows that a Bayes rule for complex conditional probability $p(m)\rightarrow p'(m)\equiv p(m|a,b)$ nolonger quantifies a change in the probability distribution (or state of knowledge) of a system when new information about the system is learned.

One potential objection is that perhaps each $p_K(a,b)$ takes values on its own curve $c_{\rho}'$, but if that were the case then no two distributions $p_K(a,b)$ and $p_K(a,b)'$ could be compared and the function $p_K(a,b)$ loses its universality which is another component of Coxian probability theory \cite{Cox,Jaynesbook,book}. It has has been shown \cite{Johansen} that the elements of a density matrix $\hat{\rho}=\sum_{a,b}\rho_{ab}\ket{a}\bra{b}$ where $\rho_{ab}=\frac{p_K(a,b)}{\braket{a}{b}}$ can be known if one measures $p_K(a,b)$, but it is, in-fact, standard probability theory (in Quantum Mechanics using weak measurements and Weak Values) which allows one to make inferences about the most probable values of $p_K(a,b)$ and thereby infer $\rho_{ab}$ - their values are inferred through measurement of spacial and angular (probability) distributions of coherent light \cite{Bollen}.
\section{\label{sec:level4} Conclusion}
It has been shown that probability distributions which take complex values are only consistent with probability theory if the distribution takes values on a nonintersecting curve $c'$ in the complex plane which can be mapped to the real number line by an order preserving operation $\mathcal{O}$ to preserve transitivity such that probability functionally ranks degrees plausibility. Because Kirkwood distributions are not consistent with Coxian probability theory, we offer the notation for the aforementioned conditional complex probability $p_K(m|a,b)\rightarrow K(m|a,b)$ (possibly better named a conditioned Kirkwood distribution or simply the Weak Value of $m$) and $p_K(a,b)\rightarrow K(a,b)$ such that they are not misinterpreted and utilized as probability distributions which rank degrees of plausibility. This development extends the extent which Kirkwood distributions differ from probability distributions; not only do Kirkwood distributions relax the first Kolmogorov axiom of probability they also are not able to rank plausibility.

\end{document}